\documentclass[twocolumn,prb,showpacs,amsmath,amssymb]{revtex4}
\usepackage{graphicx}
\usepackage{epsfig}
\usepackage{dcolumn}
\usepackage{bm}

\begin{document}
\title{Electron-phonon coupling and phonon self-energy in MgB$_2$:\\
do we really understand MgB$_2$ Raman spectra ?}
\author{Matteo Calandra}
\address{Laboratoire de Min\'eralogie-Cristallographie, case 115, 4 Place Jussieu, 75252, Paris cedex 05, France}
\author{ Francesco Mauri}
\address{Laboratoire de Min\'eralogie-Cristallographie, case 115, 4 Place Jussieu, 75252, Paris cedex 05, France}

\date{\today}

\begin{abstract}
We consider a model Hamiltonian fitted on the {\it ab-initio} band structure 
to describe the electron-phonon coupling between
the electronic $\sigma-$bands and the phonon E$_{2g}$ mode in MgB$_2$. 
The model allows for
analytical calculations and numerical treatments using very large k-point
grids. We calculate the phonon self-energy of the E$_{2g}$ mode along
two high symmetry directions in the Brillouin zone. 
We demonstrate that the contribution of the $\sigma$ bands
to the Raman linewidth of the E$_{2g}$ mode via the electron-phonon coupling
is zero. As a consequence the large resonance seen in Raman experiments 
cannot be interpreted as originated from the $E_{2g}$ mode at $\Gamma$.
We examine in details the effects of Fermi surface singularities 
in the phonon spectrum and linewidth and
we determine the magnitude of finite temperature effects in the 
the phonon self-energy. 
From our findings we suggest several possible effects which might be responsible
for the MgB$_2$ Raman spectra. 

\end{abstract}

\maketitle
\section{Introduction}
The knowledge of the MgB$_2$ \cite{Nagamatsu} electronic structure 
allows us to obtain a qualitative understanding of several peculiar
features of this material . A crucial role
is played by the $\sigma-$bands\cite{An,Belashchenko,Kortus}, formed by the in-plane 
boron-boron $sp$ bonding. Due to the small interlayer coupling
between the boron layers, these bands have a two dimensional 
character and are weakly dispersing along the 
$\Gamma $A direction.
Their corresponding Fermi surface sheets\cite{Kortus} are two slightly 
warped cylinders, with axis perpendicular to the Boron layers.
This peculiar topology results in a large contribution to the real
and imaginary parts of the phonon self energy of the E$_{2g}$ phonon mode, 
an in plane displacement of the boron atoms. 

The large contribution to the real part of the phonon self-energy has
spectacular consequences on the phonon spectrum:
the phonon frequencies of the E$_{2g}$ modes
undergo a reduction of roughly $20$ meV 
along the $\Gamma$A directions as 
predicted by {\it ab-initio} calculations\cite{Kong,Bohnen,Shukla} and
measured by high energy inelastic X-ray scattering\cite{Shukla,Baron}. 
Density functional theory  calculations of phonon\cite{Kong,Shukla,Bohnen} 
spectra indicate that the softening of the E$_{2g}$ phonon
frequencies when approaching the $\Gamma$A direction, even if
strong in magnitude, is not as abrupt as
would be expected\cite{An2} for a pure two dimensional system having
a Kohn anomaly. 
The experimental phonon dispersion\cite{Shukla} along AL and $\Gamma$M
\cite{Shukla,Baron} confirms that the E$_{2g}$ 
phonon frequencies decrease gradually as the $\Gamma$A direction is approached
and the softening at momenta corresponding to the cylinders $2k_F$ is very
small. 
The Kohn anomaly might indeed be mitigated by the presence
of a $k_z$ band dispersion and a finite temperature.
In this paper we investigate the magnitude of these two effects and discuss
their relevance in the interpretation of experimental data.

A large contribution due to the electron-phonon coupling 
is also associated with the imaginary 
part of the E$_{2g}$ phonon self-energy, the phonon linewidth.
Inelastic X-ray scattering experiments and theoretical
calculations\cite{Shukla,Baron} show an anomalously large
broadening ($\sim$ 20-30 meV FWMH) of the E$_{2g}$ mode 
along the $\Gamma$A direction only. According to ref. \cite{Baron}
the broadening of this mode is almost temperature independent, but the 
spectra displayed in fig. 4 of ref. \cite{Baron} do not allow for
a definitive conclusion since the E$_{2g}$ mode has a very small structure
factor and it is seen only as a shoulder of the close E$_{1u}$ mode.

Raman data show a completely different behaviour.
Raman experiments probe excitations
at small momentum transfer, close to the $\Gamma$ point 
of the material. The maximum momentum transfer is q$_{exp}= 2$q$_{light}$
where q$_{light}=\frac{2\pi}{\lambda}$ and $\lambda$ is the wavelength of the
incident light.
Most of the experiments are performed
with a 514.5 nm (2.41 eV) argon laser \cite{Quilty} which
corresponds to $q_{exp}=1.3*10^{-3}{\rm a_0}^{-1}\approx 0.002 \Gamma M $, 
($a_0=0.5292\AA$ is the Bohr radius).
This region  is inaccessible 
to X-ray measurement and as a consequence a direct experimental 
comparison between the two techniques cannot be performed.
Nevertheless it is instructive to compare Raman spectra with the X-ray
data as close as possible to the $\Gamma$ point.

The Raman linewidth of the E$_{2g}$ mode 
\cite{Quilty,Rafailov,Hlinka,Martinho,Goncharov,Kunc,Chen}
shows a very strong temperature dependence since
it is $\sim 20$ meV (FWMH) at 40 K and 
reaches almost 40 meV at room temperature, a factor of two larger than the one
detected in inelastic X-ray data along the $\Gamma$A direction.
Since, according to the calculation performed
in \cite{Shukla}, the anharmonic broadening at room temperature is 1.2 meV,
it cannot be responsible nor of the large linewidth neither of its strong 
temperature dependence.  

An unexplored cause of such a large temperature dependence of the linewidth 
might be the electron-phonon coupling. The electron-phonon coupling
contribution to the phonon linewidth  is indeed 
temperature dependent, (see eq. \ref{ImPI}, this work). 
Nevertheless the dependence on temperature is usually assumed to be negligible,
but no detailed studies have been performed on the subject. 

In this work we carefully analyze all the approximations involved in
the calculation of the phonon linewidth due to the electron-phonon coupling.
We analyze the temperature dependence of the
phonon linewidth and the effects of neglecting the phonon frequency in
Allen formula.
It would be highly desirable to estimate the magnitude of these
approximations using {\it ab-initio}
calculation, but the task is almost prohibitive. 
In actual ab-initio calculations\cite{Shukla,Kong} a finite number of k-points
is used together with a $\sim 0.025$ Rydberg ($\sim 3000 K$) 
smearing of the Fermi surface\cite{Degironcsmear}. Physical effects involving temperature
difference between 40 and 300 K are basically invisible to the calculation, 
since grids having at least $1000$ times larger number of k-points would be needed
\cite{footnote}. 
In this case, the calculated electron-phonon coupling
contributions and its temperature dependence in the indicated region
would be masked by computational details. 
The convergence of {\it ab-initio} calculations with the number of
symmetry-irreducible k-points is particularly relevant
for MgB$_2$\cite{Mazin}, since only the weak warping of
the two cylindrical $\sigma$ bands Fermi surface sheets  
prevents the linewidth from diverging. Moreover,
if there were effects such as anomalies in phonon spectra 
generated by the $2k_F$ singularities \cite{Baron}, they could be detected only
using a very large number of k-points in the phonon frequencies calculations.
As a consequence the use of a too small k-points mesh might affect the
calculation of  both the real and imaginary part of the phonon self-energy.

For these reasons, in this work we study the behaviour of the phonon 
self-energy of the E$_{2g}$ mode due to the electron-phonon 
interaction between this mode and the $\sigma-$bands using a model
Hamiltonian.  The Hamiltonian is composed by the two $\sigma-$bands coupled
to an harmonic dispersionless E$_{2g}$ phonon mode. The form considered for the
$\sigma-$bands is fitted from {\it ab-initio} calculations \cite{Kong} 
in the region close to the $\Gamma$ point. The phonon frequency is that of the
E$_{2g}$ at $\Gamma$. The model is illustrated in detail in sec. \ref{sec:model},
together with the form of the phonon self-energy in its real (phonon shift)
and imaginary (phonon linewidth) parts. 
The simplified form of the model allows to calculate analytically the
linewidth as ${\bf q}\to \Gamma$ along any high symmetry direction. 
Moreover it allows numerical calculations using grids of $N_k=300^3$,
symmetry-irreducible k-points in any point of the
Brillouin zone, which are enough to see temperature effects in the phonon 
linewidth due to the electron-phonon coupling.

In sec. \ref{sec:Raman}  we calculate the phonon linewidth {\it exactly} in the
limit ${\bf q}\to\Gamma$ both in its intraband and in its
$\sigma-\sigma$ interband contributions. 
We consider two cases, (i) ${\bf q}$ along the 
$\Gamma$A direction (sec. \ref{sec:numgammaa}) 
and (ii) ${\bf q}$ along $\Gamma$M, or generally along any 
direction in the $(k_x,k_y)$ plane (sec. \ref{sec:analplane}), 
since the bands and the considered
coupling are isotropic in the $(k_x,k_y)$ plane. We discuss the relevance 
of the results for the interpretation of Raman spectra.

In sec. \ref{sec:Allen} we follow ref. \cite{Allen} and derive Allen formula 
starting from the phonon self-energy, 
paying particular attention at the approximations involved. Subsequently in
sec. \ref{sec:numerical} we numerically evaluate the phonon self-energy along
the $\Gamma$A and $\Gamma$M directions for the case of a k-independent 
electron-phonon coupling. We estimate the effects of temperature and
of $\sigma-\sigma$ interband transitions. We evaluate the magnitude of the 
different approximations involved in the derivation of Allen formula.
The numerical results are also used as benchmarks to judge the reliability 
of preceding {\it ab-initio} calculations\cite{Shukla} in what concerns
the number of $k-$points used in the simulations and the value of the
smearing parameter.

Finally we question the attribution of the 77 meV peak to the
E$_{2g}$ mode and we suggest other interpretation of the Raman experiments.

 
\section{\label{sec:model}Model}

Following reference \cite{Kong}, the structure of the $\sigma$ bands close 
to the Fermi energy can be expressed as:
\begin{equation}
\epsilon_{{\bf k}n}=\epsilon_0-2t_{\perp} \cos(k_z c)-\frac{k_x^2+k_y^2}{m_{n}} {\cal R}y
\label{eq:sigmabands}
\end{equation}
where the index $n$ label the heavy/light hole bands. The holes masses are 
$m_1=0.59$ (heavy holes), $m_2=0.28$ (light holes). The average
energy along $\Gamma$A, $t_{\perp}=0.094$ eV, gives the dispersion 
of the bands. The top of the $\sigma$ bands is $\epsilon_0=0.58$ eV.
Note that $k$ are expressed in atomic units 
and $\frac{k^{2}}{m}{\cal R}y$ in eV with ${\cal R}y=13.605$.
The bands are measured respect to the Fermi energy. 
\begin{figure}[t]
\includegraphics[width=8.5cm]{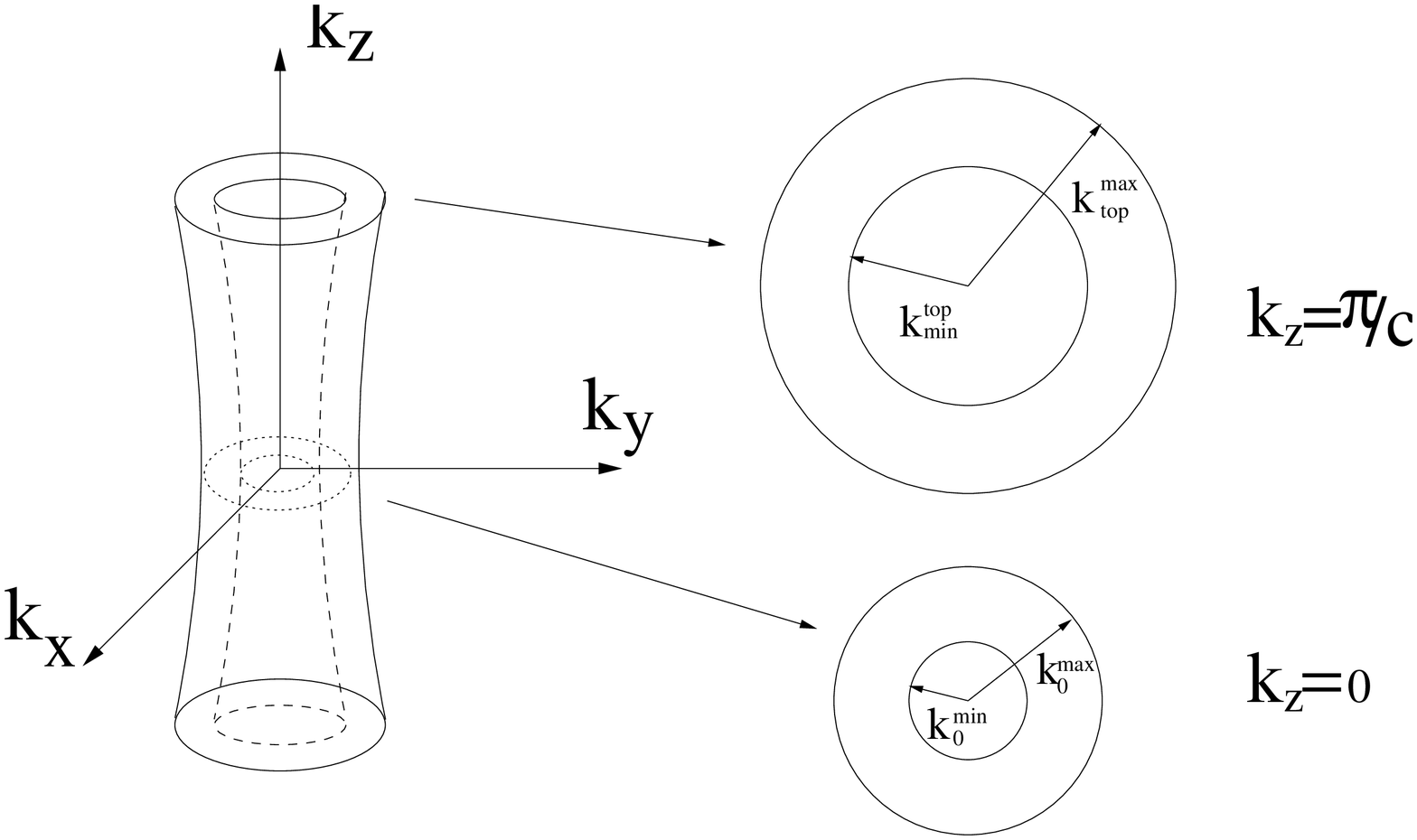}
\caption{$\sigma-$bands Fermi surface cylinders with projection over
the $k_z=\pi/c$ and $k_z=0$ planes. }
\label{fig:cylproj}
\end{figure}
The Fermi surface sheets 
identified by the bands in eq. \ref{eq:sigmabands} are
two warped cylinders (see fig. \ref{fig:cylproj}). The radii in the $k_z=0$ plane are 
$k_{0}^{\rm max}=0.13\, a_0^{-1}$ and $k_{0}^{\rm min}=0.09\, a_0^{-1}$. The
radii in the $k_z=\pm \pi/c$ planes are $k_{\rm top}^{\rm max}=0.18\, a_0^{-1} $ and 
$k_{\rm top}^{\rm min}=0.126\, a_0^{-1} $. 

The contribution to the $\nu$ phonon mode phonon self-energy due to 
the electron phonon coupling can be written as,
\begin{equation}
\Pi_{\nu}({\bf q},\omega_{{\bf q}\nu})=\frac{2}{N_k}\sum_{{\bf k},m,n} |g_{{\bf k}n,{\bf k+q}m}^{\nu}|^2\,
\frac{f_{{\bf k}+{\bf q}m} - f_{{\bf k}n}}{\epsilon_{{\bf k}+{\bf q}m}-\epsilon_{{\bf k}n}-\omega_{{\bf q}\nu}-i\eta} 
\label{eq:PIdef}
\end{equation}
where $N_k$ is the number of k-points, the sum is over the Brillouin zone and
$f_{{\bf k}n}$ are the Fermi distribution functions. 
The matrix element is 
$g_{{\bf k}n,{\bf k+q}m}^{\nu}= \langle {\bf k}n|\delta V/\delta u_{{\bf q}\nu} |{\bf k+q} m\rangle /\sqrt{2 \omega_{{\bf q}\nu}}$ where $u_{{\bf q}\nu}$ is the amplitude of the displacement of the phonon $\nu$
of wavevector ${\bf q}$,
$\omega_{{\bf q}\nu}$ its phonon frequency and $V$ the electron-ion
interacting potential\cite{footnote2}.

The real part of the phonon self-energy is
\begin{equation}
\frac{\Delta_q}{2} = \frac{2}{N_k}\sum_{{\bf k},m,n} |g_{{\bf k}n,{\bf k+q}m}^{\nu}|^2\,
{\cal P}\left[\frac{f_{{\bf k}+{\bf q}m} - f_{{\bf k}n}}{\epsilon_{{\bf k}+{\bf q}m}-\epsilon_{{\bf k}n}-\omega_{{\bf q}\nu}} \right]
\label{RePI}
\end{equation} 
where {\cal P} stands for the principal value.
With this definition, $\Delta_q$ express the renormalization of the harmonic phonon
frequencies due to electron-phonon coupling effects. 

The phonon linewidth (FWMH) is twice the imaginary part of $\Pi_{\nu}({\bf q},\omega_{{\bf q}\nu})$
divided by $N_k$, as it can also be inferred from Fermi golden rule:
\begin{eqnarray}
\gamma_{{\bf q}\nu}&=&\frac{4\pi}{N_k}
\sum_{{\bf k},m,n}|g_{{\bf k}n,{\bf k+q}m}^{\nu}|^2 \cdot \nonumber \\
&\cdot&\left(f_{{\bf k}n} - f_{{\bf k}+{\bf q}m}\right)
 \delta(\epsilon_{{\bf k}+{\bf q}m}-\epsilon_{{\bf k}n}-\omega_{{\bf q}\nu})
\label{ImPI} 
\end{eqnarray} 

In the following two subsections we calculate the phonon linewidth in the limit 
${\bf q}\to 0$ analytically. We use formula eq. \ref{ImPI} choosing {\bf q} along
two high symmetry directions in the Brillouin zone: (i) the out of plane $\Gamma$A
directions and (ii) the in-plane $\Gamma$M direction. We show that in both cases
the phonon linewidth vanishes in the ${\bf q}\to 0$ limit.   

\section{\label{sec:Raman}MgB$_2$ Raman linewidth.}

\subsection{\label{sec:numgammaa}$\Gamma$A direction}

We choose ${\bf q}$ along the $\Gamma$A direction and consider the limit for ${\bf q}$
going to zero:
\begin{equation}
\epsilon_{{\bf k}+{\bf q}m}=\epsilon_{{\bf k}n}-\frac{k_{\parallel}^2}{m_n}
\left(\frac{m_n}{m_m}-1\right){\cal R}y+2t_{\perp} qc\sin(k_z c)
\label{eq:bandexp}
\end{equation}
where $k_{\parallel}^2 = k_x^2 + k_y^2$. Eq. (\ref{eq:bandexp}) is correct at
order ${\cal O}(q^2)$.
Choosing $\omega_{{\bf q}\nu}=65$ meV (i.e. the harmonic E$_{2g}$ phonon frequency at
$\Gamma$) and substituting in eq. (\ref{eq:PIdef}) we get:
\begin{eqnarray}
\gamma_{{\bf q}\nu}=\frac{4\pi}{N_k}\sum_{{\bf k} m,n}|g_{{\bf k}n,{\bf k+q}m}^{\nu}|^2
\left(f_{{\bf k+q}m} - f_{{\bf k}n}\right)\cdot \nonumber \\
\cdot\,
\delta\left[\frac{k_{\parallel}^2}{m_n} \left(\frac{m_n}{m_m}-1\right){\cal R}y
-2t_{\perp} qc\sin(k_z c)+\omega_{{\bf q}\nu}\right]
\label{eq:gamMgB2}
\end{eqnarray}
This sum can be divided in two different contributions, one coming from 
intraband transitions ($m=n$) and the other from interband 
transitions ($m\ne n$). 

If $n=m$ and in the limit of ${\bf q}\to 0$, in eq. (\ref{eq:gamMgB2}) in order
for the $\delta$ function to be satisfied we must have that  
$\omega_{{\bf q}\nu}=2 t_{\perp} q c \sin(k_z c)$. 
The momentum used in Raman experiment \cite{Quilty}
$q_{exp}=1.3*10^{-3} {\rm a_0}^{-1}$. 
Raman scattering then samples a sphere in momentum space
centered at $\Gamma$ and of radius $q_{exp}$. It follows then
that 
\begin{equation}
|2 t_{\perp} q_{exp} c| = 1.6\, {\rm meV} \ll 
\omega_{{\bf q}\nu} =65\, {\rm meV}
\label{eq:intracond}
\end{equation}
using $c=6.653 a_0$. Thus the $\delta$-function condition in eq.\ref{eq:gamMgB2} 
is never fulfilled in Raman experiments.
The contribution to the linewidth due to intraband transition is then
exactly zero. The general fact that an optical phonon
mode cannot couple with electrons at $\Gamma$ as long as only intraband transitions are 
allowed has been already noted in the footnote number 18 of ref. \cite{Liu}.

Choosing a finite $q$ along $\Gamma$A and using eq. (\ref{eq:intracond}) we can 
determine the values of $q$ in the Brillouin zone for which the intraband contribution
is non zero, namely 
\begin{equation}
q\geq q^{\rm intra}=\frac{\omega_{{\bf q}\nu}}{2 t_{perp}c}\approx 0.052\,{\rm a_0}^{-1} \approx 0.1 \Gamma A
\end{equation}
In this estimate we have used the phonon frequency of the $E_{2g}$ at $\Gamma$, being
the phonon branches fairly flat along $\Gamma$A and ${\bf q}\to 0$ . 
We have also assumed the expansion
(\ref{eq:bandexp}) at order ${\cal O}(q^2)$ to be correct.
We will show in sec. \ref{sec:numres} that this limit is indeed correct using 
numerical calculation.

We then consider the interband contributions ($m\ne n$) and ${\bf q}\to\Gamma$.
In order for the argument of the delta function in eq. (\ref{eq:gamMgB2}) 
to be satisfied we must have that
\begin{equation}
2t_{\perp} qc\sin(k_z c)-\frac{k_{\parallel}^2}{m_n} \left(\frac{m_n}{m_m}-1\right){\cal R}y=\omega_{{\bf q}\nu}\label{eq:deltacond}
\end{equation}
In the case $n=1$ and $m=2$, we have $(\frac{m_1}{m_2}-1)>0$ since $m_1>m_2$. 
As a consequence the largest value of ${\bf q}$ for which the delta function 
is non zero is determined by the condition:
\begin{equation}
|2t_{\perp} qc|<\omega_{{\bf q}\nu}
\end{equation}
which leads to the same condition as in the intraband transition case,
namely 
\begin{equation}
q \ge 0.1\Gamma A
\end{equation}

Then we consider the term with $n=2$ and $m=1$.
In order for eq. (\ref{eq:gamMgB2}) to give a finite
linewidth at $T=0 K$, the following two conditions must be
simultaneously satisfied:
(i) the states $\epsilon_{k1}$ are occupied and the states 
$\epsilon_{k2}$ empty (and vice versa) and 
(ii) the delta function in eq. \ref{eq:gamMgB2} must be satisfied.
Recalling the Fermi surface topology of the two $\sigma$ bands, the first condition 
means that the sum is limited to the region of space included between the two warped 
cylinders. This region is included between the two cylinders having axes
along $\Gamma$A and radii $k_{0}^{\rm min}$
and $k_{\rm top}^{\rm max} $ respectively.
On the other hand for the second condition to be fulfilled we must
have  
\begin{equation}
|q|\ge \frac{k_{\parallel}^2}{m_2}\frac{0.525 {\cal R}y}{2t_{\perp}c}+
\frac{\omega_{{\bf q}\nu}}{2t_{\perp}c}
\end{equation}
where we have substituted $(\frac{m_2}{m_1} -1)=-0.525$.
The constraint imposed by the two Fermi functions 
(condition (i)) allows to replace
$k_{\parallel}$ with $k_{0}^{\rm min}$ in the inequality:
\begin{equation}
|q|\ge \frac{(k_{0}^{\rm min})^2}{m_2}\frac{0.525 {\cal R}y}{2t_{\perp}c}+
\frac{\omega_{{\bf q}\nu}}{2t_{\perp}c} > \frac{\omega_{{\bf q}\nu}}{2t_{\perp}c}
\approx 0.1 \Gamma{\rm A}
\label{eq:nointer}
\end{equation}
From eq. \ref{eq:nointer} we conclude that even the term with $n=2$ and
$m=1$ in eq. \ref{eq:gamMgB2} is zero for $|q| <  0.1 \Gamma$A. 
A similar equation can be derived for the case $\epsilon_{{\bf k}1} > 0$ and 
$\epsilon_{{\bf k}+{\bf q}2} < 0$, so that intraband transition give no
contribution to the Raman linewidth.

We have shown that both the $\sigma$ bands intraband and interband contributions 
to the phonon linewidth via the electron-phonon interaction are zero for
${\bf q}$ along $\Gamma$A and 
\begin{equation}
|q| < 0.1 \Gamma{\rm A} \approx 0.052\,a_0^{-1}
\end{equation}

\subsection{\label{sec:analplane}In-plane momenta.}

We now chose ${\bf q}$ in the $k_x,k_y$ plane and consider the limit for
 ${\bf q}$ going to zero, we have that :
\begin{equation}
\epsilon_{{\bf k}+{\bf q}m}= \epsilon_{{\bf k}n}-\frac{|k_{\parallel}+q|^2}{m_m}
{\cal R}y +\frac{|k_{\parallel}|^2}{m_n}{\cal R}y
\end{equation}
where we have chosen {\bf q} along the $\Gamma$M direction.
The phonon linewidth becomes:
\begin{eqnarray}
&\gamma_{{\bf q}\nu}&=\frac{4\pi}{N_k}\sum_{{\bf k} m,n}|g_{{\bf k}n,{\bf k+q}m}^{\nu}|^2
\left(f_{{\bf k+q}m} - f_{{\bf k}n}\right)\cdot \nonumber \\
&\cdot&
\delta\left[-\frac{|{\bf k}_{\parallel}+{\bf q}|^2}{m_m}{\cal R}y 
+ \frac{|{\bf k}_{\parallel}|^2}{m_n}{\cal R}y +\omega_{{\bf q}\nu}\right]
\label{eq:gamMMgB2}
\end{eqnarray}
We neglect terms of order $q^2$. We first consider 
intraband transition only, $(m=n)$. In this case,
we obtain:
\begin{eqnarray}
\gamma_{{\bf q}\nu}^{\rm intra}&=&
\frac{4\pi}{N_k}\sum_{{\bf k} m}|g_{{\bf k}m,{\bf k+q}m}^{\nu}|^2
\left(f_{{\bf k+q}m} - f_{{\bf k}m}\right)\cdot \nonumber \\
&\cdot&\delta\left(\omega_{{\bf q}\nu} - \frac{2 q k_{x}}{m_m}{\cal R}y \right)
\label{eq:gamMintra}
\end{eqnarray}
where $k_x$ is along $\Gamma$M.
In order for eq. \ref{eq:gamMintra} to give a non-zero value for $\gamma_{{\bf q}\nu}$
the $\delta-$function must be satisfied so that 
$2q k_{x} {\cal R}y=m_m \omega$. The
two Fermi functions limit the sum in the regions of space with (i)
$\epsilon_{{\bf k}m} < 0$ and $\epsilon_{{\bf k}+{\bf q} m}>0$ and (ii) 
$\epsilon_{{\bf k}m} > 0$ and $\epsilon_{{\bf k}+{\bf q} m}<0$. In case (i)
the sum over {\bf k} is limited to the region included by one of the two cylinders, 
depending on the value of the index $m$. The maximum k possible is 
$k_{\rm max}^{(1)}=k_{\rm top}^{\rm max}$ for $m=1$ and  
$k_{\rm max}^{(2)}=k_{\rm top}^{\rm min}$ for $m=2$.
We can then substitute $k_{\rm max}^{(1)}$ and $k_{\rm max}^{(2)}$ 
in the $\delta-$function condition in eq. \ref{eq:gamMintra} to obtain
$q_1 = 7.8\times 10^{-3} a_0^{-1}\approx 0.0125 \Gamma$M 
for $m=1$ and
$q_2 = 5.3\times 10^{-3} a_0^{-1}\approx 0.0085\Gamma$M 
for $m=2$. 
In case (ii) the sum is
limited to the region of space outside one of the two cylinders and,
since $\epsilon_{{\bf k}+{\bf q}m} < 0$ then $q>k_{\parallel}-k_{\rm top}^{\rm max}$ for
$m=1$ and $q>k_{\parallel}-k_{\rm top}^{\rm min}$ 
for $m=2$, with $q > 0$ in both cases.
Inserting $ q=k_{\parallel}-k_{\rm top}^{\rm max}$ or 
$ q=k_{\parallel}-k_{\rm top}^{\rm min}$ in the $\delta-$function
condition and solving for $k_x$ one gets
one gets 
$q_1^{\prime}=7.4\times 10^{-3} a_0^{-1}\approx 0.012 \Gamma$M 
for $m=1$ and 
$q_2^{\prime}=5.1 \times 10^{-3} a_0^{-1}\approx 0.008 \Gamma$M 
for $m=2$.
Finally the intraband contribution vanishes completely 
for 
$|\tilde{q}|<|q|<
q_{intra}={\rm min}\{q_1,q_2,q_1^{\prime},q_2^{\prime}\}= 5.1 \times 10^{-3} a_0^{-1}$,
which is factor of $4$ larger than the exchanged momentum 
$q_{\rm exp.}=1.3\times 10^{-3} a_0^{-1}$ in Raman scattering.

Note that this conservative estimate have been obtained using the $E_{2g}$ 
phonon frequency at $\Gamma$.
In the $(k_x,k_y)$ plane the $E_{2g}$ phonon modes are not degenerate and both
have phonon frequencies which are larger than the value at $\Gamma$ \cite{Shukla}.
The use of a larger $\omega_{{\bf q}\nu}$ would lead to the vanishing of the 
phonon linewidth at a larger value of {\bf q}.

Then we consider the interband case, $(m\ne n)$. We start considering $n=1$ and
$m=2$.
In order for the Fermi function 
difference in eq. \ref{eq:gamMMgB2} to be non zero one of the following 
conditions must be satisfied: (i) $\epsilon_{k 1}>0$ and $\epsilon_{k+q 2}<0$,
(ii) $\epsilon_{k 1}<0$ and $\epsilon_{k+q 2}>0$. 
The first condition means that
$|q| > k_{0}^{\rm max}-k_{0}^{\rm min} \approx 0.04 a_0^{-1} = 0.064 \Gamma M$
(the $k_z=0$ plane is where the surfaces of the two cylinders are closer).
i.e. no contribution to the linewidth for momenta smaller than $0.064 \Gamma$M.

The second condition leads to $ k_{\parallel}<k_0^{\rm max}$ and 
$|{\bf k}_{\parallel}+{\bf q}|>k_{0}^{\rm min}$. 
Thus we have:
\begin{equation}
\frac{|k_{\parallel}+q|^2 }{m_2}{\cal R}y > \frac{(k_0^{\rm min})^2} {m_2}{\cal R}y > 
\frac{(k_0^{max})^2 }{m_1}{\cal R}y
\end{equation}
meaning that the $\delta-$ function condition:
\begin{equation}
\frac{|k_{\parallel}+q|^2}{m_2}{\cal R}y +\omega_{\bf q}
=\frac{|k_{\parallel}|^2}{m_1}{\cal R}y 
\label{eq:delta2}
\end{equation}
is never satisfied.
Then we consider the case $n=2$, $m=1$. 
The Fermi functions in eq. \ref{eq:gamMMgB2}
give the following two conditions:
(i) $\epsilon_{{\bf k}2} < 0$ and $\epsilon_{{\bf k}+{\bf q}1} > 0$,
(ii) $\epsilon_{{\bf k}2} > 0$ and $\epsilon_{{\bf k}+{\bf q}1} < 0$.
In case (i) we have that $k_{\parallel} < k_{\rm 0}^{\rm min}$ and 
$|{\bf k}+{\bf q}| > k_{0}^{\rm max}$ and we get the same result of
$n=1$, $m=2$ case (i).
In case (ii) we have $k_{\parallel} > k_{\rm 0}^{min}$ and 
$|{\bf k}+{\bf q}|<k_0^{\rm max}$. We have:
\begin{equation}
\frac{k_{\parallel}^2}{m_2}{\cal R}y  > \frac{(k_0^{min})^2}{m_2}{\cal R}y  > 
\frac{k_0^{\rm max}}{m_1}{\cal R}y 
\end{equation}
and similarly to the case with $n=1$ and $m=2$, the condition \ref{eq:delta2}
is never satisfied.

In this subsection we have demonstrated that for ${\bf q}$ in the 
$k_x,k_y$ plane the linewidth vanishes at small momenta, the
intraband contribution vanishes for $|q|< 0.008 \Gamma$M 
while the interband contribution vanishes for $|q|< 0.06 \Gamma$M.
The phonon linewidth along $\Gamma$M vanishes for 
\begin{equation}
|q|< 0.008 \Gamma M \approx 4 q_{exp}
\end{equation}

\subsection{Conclusions}
 
In sections \ref{sec:analplane} and \ref{sec:numgammaa} we
have shown that the phonon linewidth due to the electron-phonon
coupling is zero in an ellipsoid centered at $\Gamma$ and having axes 
$q_{\parallel}=0.005 a_0^{-1}\approx 4 q_{exp}$ 
in the $(k_x,k_y)$ plane and $q_{\perp}=0.052 a_0^{-1}\approx 40 q_{exp}$ 
along $\Gamma$A.
Both axes are larger than the largest momenta accessible with Raman
scattering, $q_{\rm exp}=1.3\times 10^{-3} a_0^{-1}$. Along $\Gamma$A the
minimal momentum giving a final linewidth is an order of magnitude
larger than $q_{\rm exp}$.Thus if
the Raman experiment is prepared with a geometry consistent with
an exchanged momentum along the $\Gamma$A direction one should indeed
find a zero linewidth for the $E_{2g}$ mode. Although this 
geometry is currently employed in most of the Raman experiments
in MgB$_2$, it seems that a large linewidth ($\approx 40 meV$) is
detected in the Raman data published up to now. We therefore conclude
that the broad feature visible in these experiments cannot be associated
to a pure $E_{2g}$ phonon excitations whose linewidth is determined
by the electron-phonon coupling of the $E_{2g}$ at $q_{\rm exp}$. 
In the final section of the paper we put
forward possible explanations for the experimental spectra.

\section{\label{sec:Allen}Allen formula}

The linewidth $\gamma_{{\bf q}\nu}$ can be related to the electron-phonon coupling
\cite{Allen} via a simple approximations. 
Namely, at temperature such that $k_bT\gg \omega_{{\bf q}\nu}$ or in the
case of a temperature independent $\gamma_{{\bf q}\nu}$, using the
$\delta$-function condition 
$\delta(\epsilon_{{\bf k}+{\bf q}m}-\epsilon_{{\bf k}n}-\omega_{{\bf q}\nu})$ 
in eq. (\ref{ImPI})  one can substitute in formula (\ref{ImPI}) 
\begin{equation}
\omega_{{\bf q}\nu} \, \frac{f_{{\bf k+q}m}-f_{{\bf k}n}}{\omega_{{\bf q}\nu}}\longmapsto \omega_{{\bf q}\nu}\,\left.\frac{\partial f}{\partial \epsilon}\right|_{\epsilon=\epsilon_{{\bf k}n}}
\label{eq:subsdelta}
\end{equation}
If the temperature dependence in equation (\ref{ImPI}) is weak than the Fermi function 
can be considered a step function, so that:
\begin{equation}
\tilde{\gamma}_{{\bf q}\nu} = \frac{4\pi\omega}{N_k}\sum_{{\bf k},m,n}|g_{{\bf k}n,{\bf k+q}m}^{\nu}|^2\delta(\epsilon_{{\bf k}n})\delta(\epsilon_{{\bf k}+{\bf q}m}-\epsilon_{{\bf k}n}-\omega)
\label{gammadeldel}
\end{equation} 
If $T_0$ is the highest temperature
for which the substitution of the derivative of the Fermi function 
with a Dirac $\delta-$function in eq. \ref{gammadeldel} is still correct,
then eq. \ref{gammadeldel} is  valid in the range of temperatures such that
$k_bT_0 > k_bT\gg \omega_{{\bf q}\nu}$. If $\gamma_{{\bf q}\nu}$ is
temperature independent, then the condition is simply $T < T_0$.
It is worth noting that in the limiting case of a very large phonon frequency it might 
occur that $k_bT_0 < \omega_{{\bf q}\nu}$ 
and formula \ref{gammadeldel} might be never valid. 
Since in practice one has phonon frequencies which are of the order of
$300 K$ or more, the only real condition of applicability of eq. \ref{gammadeldel}
is that $\gamma_{{\bf q}\nu}$ has to be temperature independent.
 
From the definition \cite{Grimvall} of the electron-phonon coupling 
($\lambda_{{\bf q}\nu}$) for the mode $\nu$ at
 point ${\bf q}$ one sees that 
\begin{equation}
\lambda_{{\bf q}\nu}= \frac{\tilde{\gamma}_{{\bf q}\nu}}{2\pi N(0) \omega_{{\bf q}\nu}^2}
\label{eqAllen}
\end{equation}
which is Allen Formula\cite{Allen}. Allen formula allows to extract the 
electron phonon coupling from the measured linewidth under the assumption that 
anharmonic effects are negligible. For MgB$_2$ this condition is 
fulfilled along the $\Gamma$A direction \cite{Shukla}.

In actual calculations, it is customary to neglect the frequency dependence in the
$\delta$ function in eq.(\ref{gammadeldel}), obtaining
\begin{equation}
\gamma^{0}_{{\bf q}\nu} = \frac{4\pi\omega_{{\bf q}\nu}}{N_k}
\sum_{{\bf k},m,n}|g_{{\bf k}n,{\bf k+q}m}^{\nu}|^2
\delta(\epsilon_{{\bf k}n})\delta(\epsilon_{{\bf k}+{\bf q}m})
\label{gammadeldelnoom} 
\end{equation} 

This assumption is unjustified at ${\bf q}=0$ and leads to 
the wrong behaviour at $\Gamma$. 
Thus formula (\ref{gammadeldelnoom}) cannot be used 
to explain finite temperature Raman experiments due 
(i) to its (wrong) behaviour at $\Gamma$ and
(ii) to the lack of temperature dependence. The correct
behaviors are included in expression (\ref{ImPI}).

\section{\label{sec:numerical}Numerical calculations}

\subsection{Model for the electron-phonon coupling matrix element.}

We consider a model composed of the two $\sigma$ bands in eq.(\ref{eq:sigmabands})
coupled to the phonons through a k-independent coupling. The electron
phonon matrix element is: 
$g_{{\bf k} m,{\bf k}+{\bf q} n}=g\delta_{m,n}+\alpha g (1-\delta_{mn})$, 
where $m,n$ run over the two $\sigma$ bands and $\alpha$ determines the 
magnitude of the interband transitions 
($\alpha=0$ correspond to the case where interband transition are suppressed).
We assume only one dispersionless phonon mode whose phonon frequency is determined from 
the calculated E$_{2g}$ phonon frequency
at $\Gamma$ \cite{Shukla}, namely 
$\omega_{{\bf q}} = 65 {\rm meV} = 754 $K. 
Along the $\Gamma$A direction, as
confirmed by inelastic X-ray scattering data \cite{Shukla}, this approximation
is fairly correct for the $E_{2g}$ mode.
Considering only one phonon mode, from now on we 
drop the index $\nu$ from the linewidth definitions.

\subsection{Technical details}

In the following subsections we calculate the real (eq. \ref{RePI}) and 
imaginary (eq. \ref{ImPI}) parts of the phonon self-energy (eq. \ref{eq:PIdef})
in the $k_x,k_y$ plane and along the $\Gamma$A direction.

In the calculations of the real part of the phonon self energy 
along $\Gamma$M we consider a finite temperature and we implement  
eq. \ref{eq:PIdef} with a $\eta$ smearing of $350 K$. This 
smearing is necessary to calculate the principal value in eq. \ref{eq:PIdef}.
Thus we extract the real part at the end. 
This procedure gives a faster convergence as a
function of $N_k$. The sums are performed using a grid of 
$N_k=300^2$ for the two dimensional case with $t_{\perp}=0$ 
and $N_k=300^3$ for the three dimensional case 
with $t_{\perp}\ne 0$. In both cases the grids are formed
by $N_k$ {\it symmetry-irreducible} $k-$points, obtained from a mesh centered
at $\Gamma$ and randomly displaced from the origin.

In the calculation of the imaginary part  
we replace the Dirac delta functions with Gaussians of width $\sigma$, namely:
\begin{equation}
\delta(x) \to \frac{e^{-\frac{x^2}{\sigma^2}}}{\sqrt{\pi \sigma}}
\end{equation} 
We compute then $\gamma_{{\bf q}}$, 
$\tilde{\gamma}_{{\bf q}}$ and $\gamma^{0}_{{\bf q}}$ 
(using equations \ref{ImPI},
\ref{gammadeldel} and \ref{gammadeldelnoom} respectively)
as a function of $\sigma$ on a given mesh of 
$N_k$ {\it symmetry-irreducible} k-points.
We then repeat the calculation on grids with always higher $N_k$ 
(with $N_k$ up to $300^3$) in order to perform the limits 
of $N_k \to \infty$ and $\sigma \to 0$. In this way we obtain the
continuum limit.
The comparison between the results for the phonon linewidth obtained
using the different formulas 
allows to judge the reliability of the 
different approximations in the 
calculation of the phonon linewidth.
\begin{figure}[t]
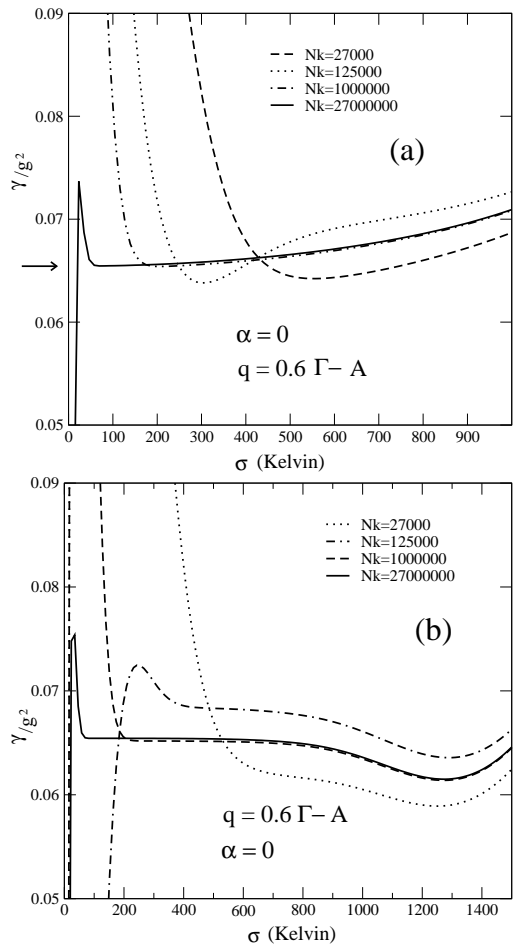

\includegraphics[width=6.7cm]{gausscaleT300q0_6.eps}
\includegraphics[width=6.7cm]{lwhermitegT300q0_6.eps}
\caption{Phonon linewidth calculated at $T=300 K$ using in formula \ref{ImPI}
as a function of the Gaussian smearing ($\sigma$), of the number 
$N_k$ of k-points used in the sum. 
The arrow indicates the value that can be extracted from the largest
mesh calculation. In (a) a pure Gaussian smearing has been used, while
in (b) an Hermite-Gaussian smearing of order 1}
\label{fig:lwscalegauss}
\end{figure}

In figure \ref{fig:lwscalegauss} (a) and (b) we show the convergence of the 
linewidths  $\gamma_{{\bf q}}$ (${\bf q}=0.6\Gamma$A) 
as a function of the Gaussian smearing $\sigma$ (expressed in Kelvin).
In (a) we used a Gaussian smearing, in (b) an hermitian-Gaussian smearing of
order 1\cite{Degironcsmear}.
As can be seen, the dependence on the smearing is fairly weak 
for the largest mesh ($N_k=27\times10^6$). 
In this case a value of $\sigma$ included between
50 and 100 K gives a result almost converged to the continuum limit.
From now on we adopt this mesh and this range of $\sigma$ values to obtain 
the continuum limit of our $\gamma$ v.s. $\sigma$ curves.

Note that previous {\it ab-initio} calculations of the phonon
linewidth \cite{Shukla} have been performed with mesh of
$27000$ symmetry-irreducible k-points. 
For the case of an Hermite Gaussian smearing, as can be seen in
the picture, this would lead to an error in the estimation 
of the phonon linewidth of the order
of $\approx 5 \%$.

\subsection{\label{sec:numres}Phonon self-energy due to the electron-phonon coupling}

\subsubsection{\label{sec:tperp}Effect of the band-dispersion along $k_z$.}

In the $(k_x,k_y)$ plane, the band structure of eq. \ref{eq:sigmabands}
is composed of two bands each of them formed by a free electron like
dispersion. As a consequence in the $t_{\perp}=0$, $\alpha=0$ and $T=0 K$ case 
(purely two dimensional with non interacting bands and at zero temperature) 
one expects to find two singularities 
(one for each band) at $2k_{F}$
in the imaginary part of the phonon self-energy
and in the first derivative of the real part of the phonon self-energy.
\begin{figure}[h]
\includegraphics[width=7.0cm]{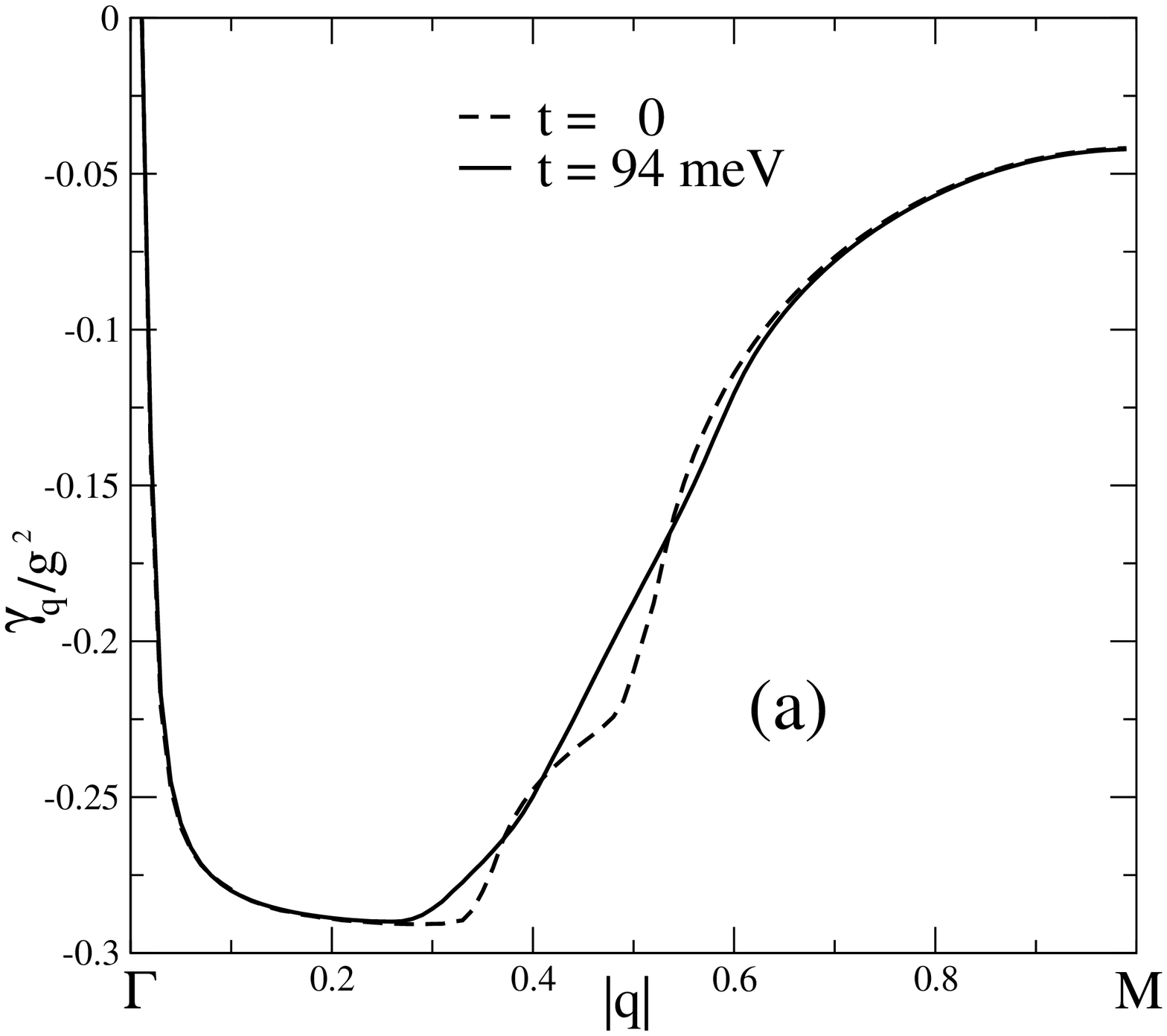}
\includegraphics[width=7.0cm]{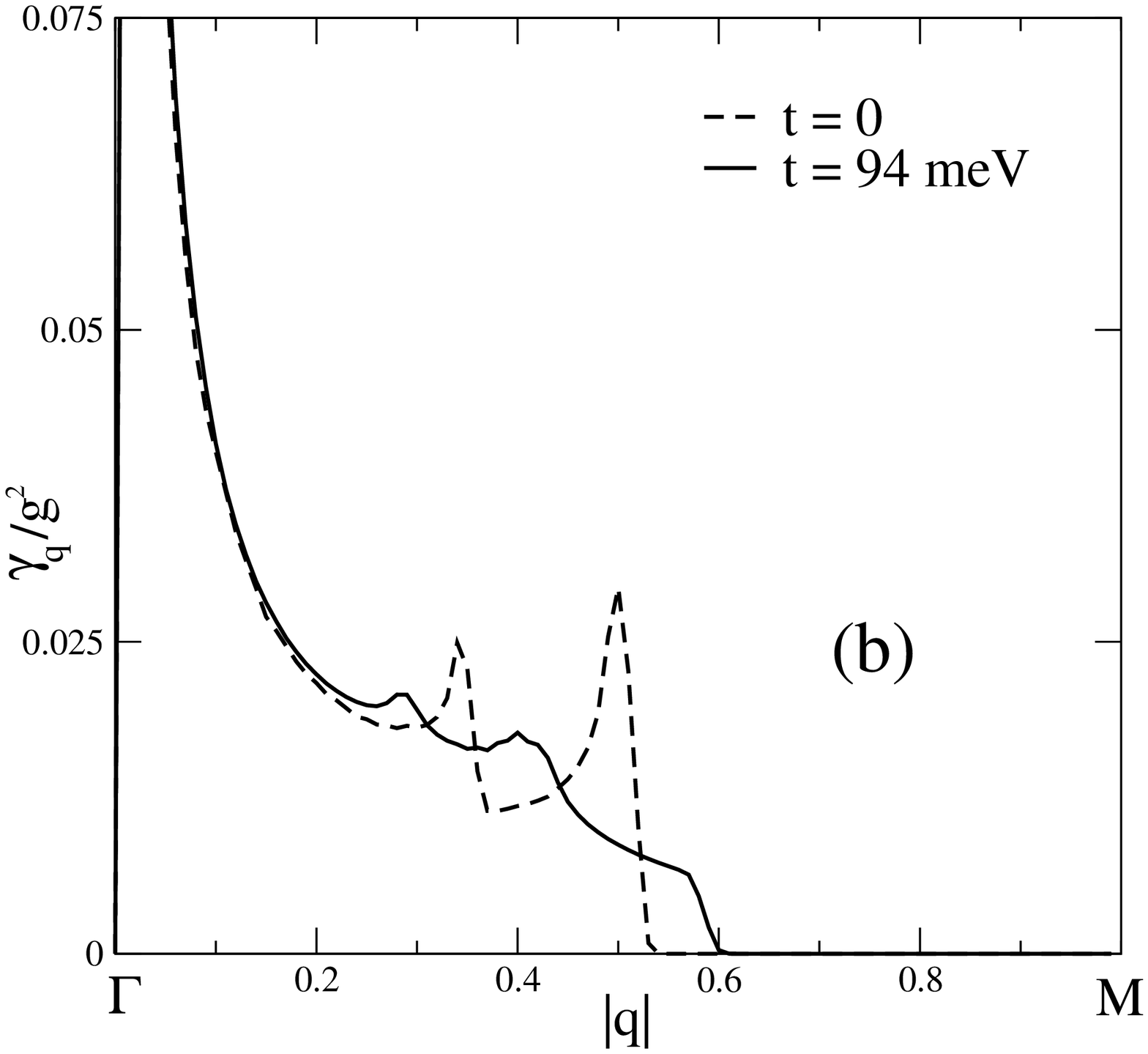}
\caption{Real (a) and imaginary (b) part of the phonon self-energy of the
$E_{2g}$ mode for $t_{\perp}=0$ (dashed lines) and $t_{\perp}=94$ meV
(continuous line) calculated for ${\bf q}$ along the $\Gamma$M direction
and at $T=40 K$. Interband transition have been suppressed.}
\label{fig:PInoint}
\end{figure}
This is shown in fig. \ref{fig:PInoint} (a) and (b) (dashed lines)
for the real and imaginary part respectively. In the real part
the singularities in the first derivative 
are seen as cusps at $2k_{F}$. 
At $T=0$ the slope on the right of each cusp should be infinite. A
finite slope is obtained as long as a finite non-zero temperature
is used \cite{Kim} (even for small temperatures the slope
at $2k_{F}$ is not vertical).
These singularities are originated by the behaviour of the response 
function in two dimensions at $T=0$ and are smoothed out
at finite temperature\cite{Kim}.
In three dimension ($t_{\perp}\ne 0$)
the singularities should disappear
\cite{Fetter,Kim}. The real part becomes continuous with no cusps and 
in the imaginary part the singularities are replaced by
 smeared continuous peaks. 
The level of smearing is determined by the three dimensional character of the 
system, in our case by the strength of $t_{\perp}$.
Since the sigma bands in MgB$_2$ have a small $t_{\perp}$
it is important to determine how far is the system from the two
dimensional case.

As can be seen in fig. \ref{fig:PInoint} (a) and (b) (continuous line)
the singularities are strongly affected even in the case of a small
$t_{\perp}$. Indeed the real part presents only very smeared cusps 
corresponding to the three-dimensional $2k_F$ positions. 
Similarly the imaginary part presents two smeared peaks at $2k_{F}$.
Even the small $t_{\perp}$ considered in this paper is sufficient
to basically eliminate the effects of the two-dimensional
$2k_{F}$ singularities.

We also study the behaviour of the phonon linewidth along 
the $\Gamma$A direction for $t_{\perp}\ne 0$. 
Along this direction, the phonon linewidth vanishes
for $|{\bf q}|<0.1 \Gamma$A, as demonstrated in sec. \ref{sec:numgammaa}. 
This is confirmed by the numerical 
calculations reported in fig. \ref{fig:gammaGA} (continuous line).
The phonon linewidth
increases monotonically approaching ${\bf q_0}=0.1 \Gamma$A from larger momenta and
becomes singular for ${\bf q}\to {\bf q_0}^+$, due to the behaviour 
caused by the $\delta-$function in eq. \ref{ImPI}.

\subsubsection{\label{sec:interband}Effect of the interband transitions between 
the $\sigma$ bands}

\begin{figure}[h]
\includegraphics[width=7cm]{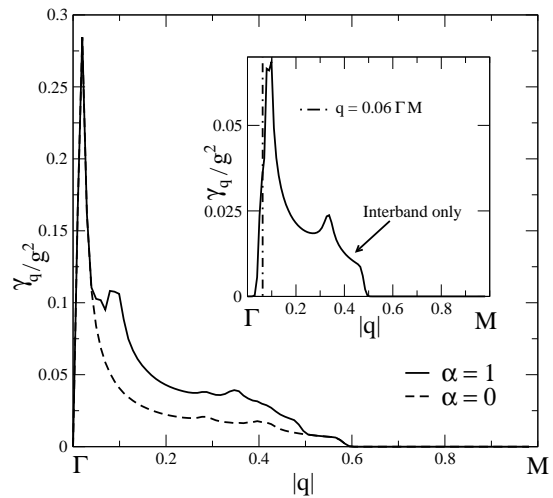}
\caption{Imaginary part of the phonon self-energy of the
$E_{2g}$ mode  calculated at $T=40 K$ 
for ${\bf q}$ along the $\Gamma M$ direction and at $T=40 K$ 
using in formula \ref{RePI} and \ref{ImPI} for the real and imaginary 
part respectively. $\alpha=0$ corresponds
to the absence of interband transition. Inset: Intraband contribution
to the $E_{2g}$ phonon linewidth. The dashed
lines is $q=0.06\Gamma$M, the limit derived analytically 
in sec. \ref{sec:analplane} for the vanishing of intraband transition.}
\label{fig:PIintra}
\end{figure}

In this section we consider the effect of interband transitions, choosing
$\alpha=0,1$ and $t_{\perp}=94 meV$. The calculated imaginary part
of the phonon self-energy at $T=40 K$ is illustrated
in fig. \ref{fig:PIintra}. Besides the $2k_F$ features 
found in the $\alpha=0$ case 
we found several other features which originates from the interband
contribution. 
The interband contribution drops to zero
at $q\approx 0.06 \Gamma$M, 
as was predicted in section \ref{sec:analplane} and as shown in the inset of
fig.  \ref{fig:PIintra}.

Along the $\Gamma$A direction the interband transitions are
completely negligible. This can be seen in fig. \ref{fig:gammaGA}
where the two curves with $\alpha=1$ and $\alpha=0$ are 
indistinguishable on the scale of the picture.

\subsubsection{\label{sec:temp}Temperature effects.}

Besides a finite dispersion along the $k_z$ axis, a second effect 
responsible for the smearing out of the singular features at $2k_F$ is
the finite temperature.
\begin{figure}[h]
\includegraphics[width=7cm]{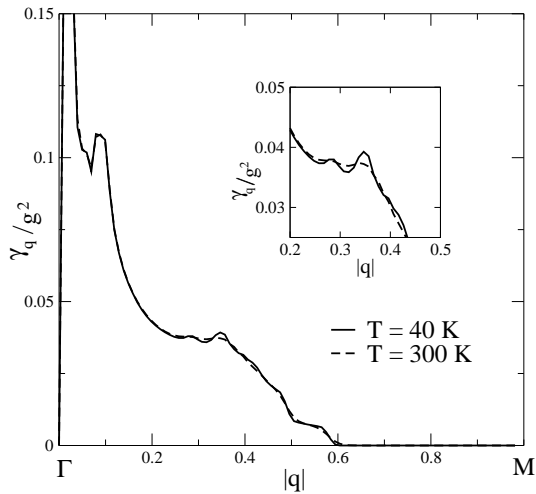}
\caption{Imaginary part of the phonon self-energy of the
$E_{2g}$ mode calculated at $T=300 K$ and $T=40 K$ for $\alpha=1$ and 
${\bf q}$ along the $\Gamma$M 
direction using eq. \ref{ImPI}. }
\label{fig:PIintraT300}
\end{figure}
In fig. \ref{fig:PIintraT300} we show the phonon linewidth
for momenta along $\Gamma$M for $T=40 K$ and $T=300K$. Overall
there is a very weak dependence on temperature.
Finite temperature effects (between 40 and 300 K) in the $(k_x,k_y)$ 
plane are larger close to the $2k_F$ singularities 
(see insect fig. \ref{fig:PIintraT300}). Nevertheless, when 
compared to the value of the phonon-linewidth, temperature effects are fairly
small and negligible in the calculations of the phonon linewidth.
Moreover, as can be seen clearly in the inset of fig. \ref{fig:PIintraT300},
the singular behaviour of the two dimensional $2k_F$ feature 
is completely lost. 

For the $\Gamma$A direction the effect is even smaller, indeed the results of the 
calculations at $T=40 K$ and $T=300 K$ are indistinguishable on the scale of
fig. \ref{fig:gammaGA} 

\begin{figure}[h]
\includegraphics[width=7cm]{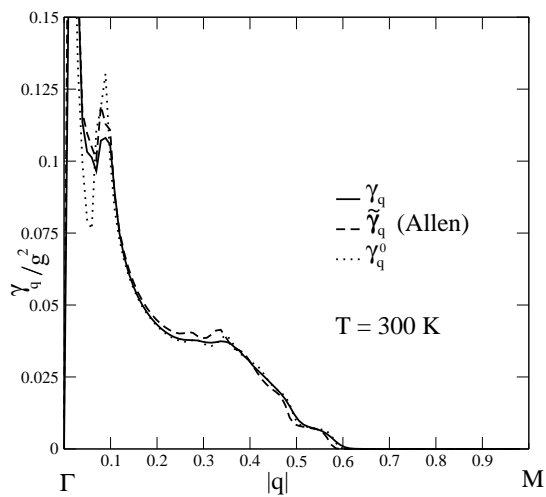}
\caption{Phonon linewidth of the
$E_{2g}$ mode  calculated at $T=300 K$ and $\alpha=1$ 
for ${\bf q}$ along the $\Gamma$M direction using 
eq.  \ref{ImPI} ( $\gamma_{\bf q}$ ), eq.\ref{gammadeldel}
( $\tilde{\gamma}_{\bf q}$ ) and eq. \ref{gammadeldelnoom} 
( $\gamma_{\bf q}^0$ ).} 
\label{fig:AllenomGM}
\end{figure}
\begin{figure}[h]
\includegraphics[width=7cm]{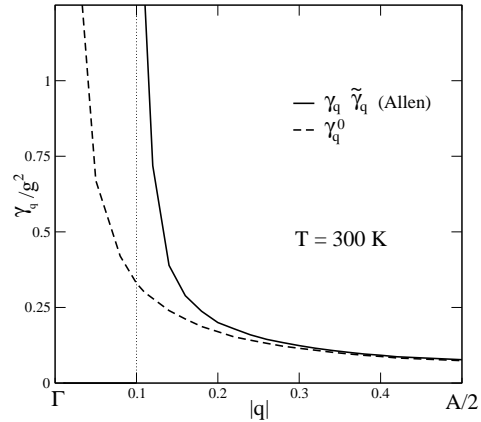}
\caption{Phonon linewidth at $T=300 K$ for $q$ along the $\Gamma$A
direction. The $\alpha=1$ case is overlapped to the $\alpha=0$,
the contribution from intraband transition is very small. The 
linewidth vanishes for $q \le 0.1 \Gamma$A (dotted line).}
\label{fig:gammaGA}
\end{figure}

\subsubsection{\label{sec:allen}Allen formula}

In fig. \ref{fig:AllenomGM} and fig. \ref{fig:gammaGA} we compare the linewidth 
calculated using $\gamma_{\bf q}$ (eq. \ref{ImPI}),
$\tilde{\gamma}_{\bf q}$ (eq. \ref{gammadeldel})
and $\gamma_{\bf q}^0$ (eq. \ref{gammadeldelnoom}) along
the $\Gamma$A direction and $\Gamma$M directions respectively. 

In passing from the $\gamma_{\bf q}$ to $\tilde{\gamma}_{\bf q}$ we have
assumed the linewidth to be temperature independent. In the preceding section 
(sec \ref{sec:temp}) we have shown that this is indeed the case, so that we
expect $\gamma_{\bf q}\approx \tilde{\gamma}_{\bf q}$ almost everywhere in the
Brillouin zone. This is what is seen in fig. \ref{fig:AllenomGM} and 
fig. \ref{fig:gammaGA}. These two pictures justify the use of Allen formula
for MgB$_2$.

From $\tilde{\gamma}_{\bf q}$, $\gamma_{\bf q}^0$ is obtained 
by neglecting the phonon frequency in
one of the two $\delta-$functions of eq. \ref{gammadeldel}.
This approximation leads to unpredictable
results which, as a consequence, must be investigated case by case, since the magnitude
of the effects produced by this approximation crucially depends on details of
the band structure close to the Fermi level and on the value of the phonon 
frequency. 

As shown in fig. \ref{fig:AllenomGM}, for ${\bf q}$ along the $\Gamma$M 
direction this approximation is fairly well justified. On the contrary along
$\Gamma$A (see fig. \ref{fig:gammaGA}) $\gamma_{\bf q}$
and $\gamma_{\bf q}^0$ display two completely different behaviors.
This is mainly due to the fact that $\gamma_{{\bf q}}$
is singular at $\Gamma$, while $\gamma_{\bf q}^0$ 
is singular for ${\bf q}\to {\bf 0}$ (see sec. \ref{sec:numgammaa}).
Moreover, as we have shown in sec. $\gamma_{\bf q}=0$ for $q < 0.1 \Gamma$A. 
The proper behaviour is recovered in the region 
$0.3 \Gamma$A $<|q| < 0.5\Gamma$A where we find that 
$\gamma_{\bf q}^0 \approx \gamma_{\bf q}$. 

\section{Conclusions}

In this work we have studied the behaviour of the phonon self-energy of the 
E$_{2g}$ mode, both in its real and imaginary part. Our conclusions can be
summarized in the following three points:
\begin{enumerate}
\item{{\it Suppression of Fermi surface singularities in phonon dispersion and 
linewidth:} two dimensional systems display 2$k_F$ singularities in the phonon
spectrum and linewidth. Naively one would expect MgB$_2$ to be similar, being
the band dispersion along the $k_z$ axis very small. On the contrary we have 
shown in sec. \ref{sec:tperp} that even such a small $t_{\perp}$ strongly suppress
the $2k_F$ singularities, so that the phonon spectrum becomes rather smooth,
and the singularities in the phonon linewidth are removed. An additional
effect (see sec. \ref{sec:temp}) is given by finite temperature which at 300 K 
completely washes out any feature in the imaginary part of the phonon self-energy.}

\item{{\it Behaviour of the phonon linewidth for ${\bf q}\to \Gamma$:} We have
shown that the phonon linewidth, both in its intraband and interband contributions
vanishes in an ellipsoid centered at $\Gamma$ and having axes 
$q_{\parallel}=0.008 \Gamma$M
in the $(k_x,k_y)$ plane and $q_{\perp}=0.1 \Gamma$A along the $k_z$ axis.
The two values are larger than the Raman momentum $q_{exp}$, namely
$q_{\parallel}\approx 4 q_{exp}$ and $q_{\perp}\approx 40 q_{\exp}$.
This calculation demonstrates that the huge linewidth seen in Raman
experiments cannot be attributed to the E$_{2g}$ mode.}

\item{{\it Temperature dependence in $\gamma_{{\bf q}\nu}$ and reliability of
Allen formula $\tilde{\gamma}_{{\bf q}\nu}$ and of $\gamma_{{\bf q}\nu}^0$:} 
the phonon linewidth is almost temperature independent in the T=0-300 K region. 
Small temperature effects are detected close to $2k_F$, but always
less than some percent of the total linewidth. 
Since the phonon linewidth is basically temperature 
independent the use of Allen formula $\tilde{\gamma}_{{\bf q}\nu}$ is justified 
in the full Brillouin zone. On the contrary the approximations customary employed
in {\it ab initio} calculations of neglecting the phonon frequency in one of
the $\delta-$functions in $\tilde{\gamma}_{{\bf q}\nu}$, obtaining 
$\gamma_{{\bf q}\nu}^0$ is not always justified. 
Along the $\Gamma-A$ directions the neglecting of the phonon frequency in
the $\delta-$function shifts
the singularity at $q\approx 0.1 \Gamma-A$ to the $\Gamma-$ point, leading to 
a completely wrong behaviour which affects all the region with $q< 0.3 \Gamma$A.}
\end{enumerate}

\section{Consequences for the interpretation of Raman spectra}

An immediate result of these three points is that the interpretation of the
broad feature seen in Raman spectra at 77 meV 
\cite{Quilty,Rafailov,Hlinka,Martinho,Goncharov,Kunc,Chen}
as a phonon excitation due to the E$_{2g}$ mode at $\Gamma$ 
is not correct. 
Indeed we have shown in this work that the huge temperature dependence of 
the Raman linewidth cannot be explained by a temperature effect 
in the electron-phonon contribution to the phonon linewidth. 
In a preceding work \cite{Lazzeri} we 
showed that the anharmonic contributions to the phonon linewidth has a weak
temperature dependence. As a consequence the temperature dependence found in 
Raman data remains completely unexplained. 
Besides its temperature dependence, the value of the Raman linewidth at $\Gamma$
is not consistent with the theoretical findings. Indeed we have demonstrated that
electron-phonon contribution to the phonon linewidth is zero in an ellipsoid
centered in $\Gamma$ and larger than the Raman exchanged momentum. 
This is not at all the case for what concerns Raman spectra.

There are additional considerations, concerning the position of the
77 meV feature, which seem to indicate that it is very unlikely that it
can be interpreted as due to a phonon excitation at $\Gamma$.
The calculated harmonic phonon frequency of the E$_{2g}$ mode 
at $\Gamma$ is indeed $65$ meV, a value 15\% smaller than the experimental result.
This has lead several groups to the conclusion that the difference might be
due to anharmonic effects\cite{Kortus,Liu,Yildirim,Choi}.  A careful 
determination\cite{Lazzeri} of the anharmonic phonon frequency shift,
explicitly taking into account three and four phonon vertexes 
and the scattering between different phonon modes
at different q-points in the Brillouin zone gives a fairly small
value of this quantity at  $\Gamma$ (+5\% of the harmonic phonon
frequency, $\approx 3.12$ meV), 
clearly to small to justify the feature at 77 meV.
This is confirmed by inelastic X-ray data of two independent groups 
\cite{Shukla,Baron} showing phonon spectra in good agreement with 
the harmonic phonon frequencies, suggesting
small anharmonic effects. Unfortunately inelastic X-ray scattering is not 
possible at the zone center, so that a direct comparison of the spectra
cannot be performed. 

In what follows we analyze several hypothesis that can be made in order 
to reconcile, theory and inelastic X-ray data with experiments with
Raman data.

A possibility is that the 77 meV feature in Raman data could be 
ascribed to a single resonant process involving the E$_{2g}$ phonon 
mode coupled to electronic
excitations. This would be consistent with the  asymmetric shape
of the peak, reminiscent of a  Fano resonance\cite{Fano}.
As a consequence the position of the peak would not correspond
to the E$_{2g}$ mode phonon frequencies but it would be slightly
shifted to lower frequencies. The temperature dependence
of the linewidth might be different in this case.
In this case it is interesting to study the peak position as
a function of the energy of the incident light.
A study of the dependence of the spectrum from the wavelength 
of the incident light has been performed in ref. \cite{Rafailov}.
It is shown that as the wavelength is changed, the peak energy position 
remains basically the same, even if the shape changes substantially. 
In the same work, from the study of the
depolarization ratio between parallel and perpendicular orientations
of the incident and emitted light, it is concluded that the symmetry 
of the excitation {\it cannot} be that of a single E$_{2g}$ mode,
supporting the idea illustrated in this paper that the Raman does not measure
the E$_{2g}$ phonon excitation at $\Gamma$.

An alternative scenario is that the Raman peak might be due to excitation 
of phonons which are not at the $\Gamma$ point. 
Such an excitation can be activated by (i) the presence
of defects such as Mg vacancies, (ii) multi-phonon scattering.
A defect breaks translational symmetry and it makes possible
to observe in Raman spectra phonon excitations at non-zero momenta.
Indeed in a similar system such as defected Graphite, due to the almost two dimensional
character of the electronic structure and a strong electron-phonon coupling
\cite{Thomsen,Piscanec}, the phonon at the K-zone boundary has a very strong signal
in Raman spectra (known as the D peak).
However a defect activated peak cannot explain alone the strong temperature
dependence of MgB$_2$ Raman linewidth between
40 and 300 K. This temperature range is not high enough
to change the population of a phonon at 77 meV. 
The temperature dependence might be explained by a multi-phonon process 
such as the absorption of an acoustic phonon and emission of an optical
phonon with opposite non-zero momenta.  
Multi-phonon scattering is also seen in graphite\cite{Cancado} and is responsible
for the G$^{\prime}$ peak observed in Raman spectra.

\section{Acknowledgments}

We acknowledge illuminating discussion with A. Shukla, M. d'Astuto and A. C. Ferrari.
The calculations were performed at the IDRIS supercomputing center (project 031202).


\begin{thebibliography}{99}
\bibitem{footnote} The convergence is given by the number of k-points in the 
symmetry-irreducible wedge of the Brillouin Zone.
\end{thebibliography}
\end{document}